\documentclass[twocolumn,preprintnumbers,nofootinbib,showpacs,showkeys]{revtex4}

\usepackage{graphicx}
\usepackage{graphics}
\usepackage{dsfont}
\usepackage{epsfig}
\usepackage{amsmath,amssymb,amsthm,amscd}

\usepackage{hyperref}
\newcommand{\beq}{\begin{equation}}
\newcommand{\eeq}{\end{equation}}
\newcommand{\be}{B_\oplus}

\newcommand{\nm}{n_{-}}
\newcommand{\np}{n_{+}}

\def\be{\begin{equation}}
\def\ee{\end{equation}}
\def\baray{\begin{eqnarray}}
\def\earay{\end{eqnarray}}
\def\bes{\begin{eqnarray}}
\def\ens{\end{eqnarray}}
\def\ba{\begin{eqnarray}}
\def\ea{\end{eqnarray}}

\def\bes {\begin{eqnarray}}
\def \ens {\end{eqnarray}}

\def\Tr{\textrm{Tr}}

\begin{document}

\newcommand{\bea}{\begin{eqnarray}}
\newcommand{\eea}{\end{eqnarray}}
\newcommand{\barr}{\begin{array}}
\newcommand{\earr}{\end{array}}

% page numbers bottom-center
\pagestyle{plain}

\preprint{USTC-ICTS-18-24}

\title{Recent Developments in Topological String Theory}

\author{Min-xin Huang\footnote{email: minxin@ustc.edu.cn} 
}

\affiliation{Interdisciplinary Center for Theoretical Study, School of Physical Sciences \\     University of Science and Technology of China,  Hefei, Anhui 230026, China
}

%%%%%%%%%%%%%%%%%%%%%%%%%%%%%%%%%%%%%%%%%%%%%%%%%%%%%%%%%%%%%%%%%%%%%%%%%%%%

\begin{abstract}
This review summarizes the recent developments in topological string theory from the author's perspective, mostly focused on aspects of research in which the author is involved.  After a brief overview of the theory, we discuss two aspects of these developments. First, we discuss the computational progress in the topological string partition functions on a class of elliptic Calabi-Yau manifolds. We propose to use Jacobi forms as an ansatz for the partition function. For non-compact models, the techniques often provide complete solutions, while for compact models, though it is still not completely solvable, we compute to higher genus than previous works. Second, we explore a remarkable connection of refined topological strings on a class of non-compact toric Calabi-Yau threefolds with non-perturbative effects in quantum-mechanical systems. The connections provide rarely available exact quantization conditions for quantum systems and new insights on non-perturbative formulations of topological string theory. 

\end{abstract}


\maketitle

\tableofcontents

\section{Introduction}\label{section1}

In the mid-1980's, physicists realized that string theory, which was originally formulated as a theory of strong interactions, is a promising candidate for reconciling quantum mechanics with gravity \cite{Green:1984sg}. When constructing four-dimensional $\mathcal{N}=1$ supersymmetric gauge theories that resemble our world, the most suitable option at the time was the compactification of $E_8\times E_8$ heterotic string theory on Calabi-Yau three-folds \cite{Gross:1984dd, Candelas:1985en}. Subsequently, the properties of Calabi-Yau manifolds have been studied intensely.

Although a convincing quantitative verification of string models using accepted parameters in the Standard Model of particle physics is out of reach, theoretical studies of Calabi-Yau manifolds have provided surprising deep insights at the boundary between mathematics and physics. These insights have inspired the formulation of topological string theory and the discovery of mirror symmetry. Some of the earliest pioneering works on this subject are \cite{Witten:1988xj, Bershadsky:1993cx, Greene:1990ud, Candelas:1990rm}. Since this brief article cannot provide a comprehensive review of topological string theory, which is by now a rather big subject, we will barely give a brief introduction as a prelude to the recent developments.  See e.g. the book \cite{MirrorSymmetry} for a comprehensive review.

String theory is usually formulated as a sigma model on a two-dimensional worldsheet, where bosonic fields are the coordinates of a target space. The main idea of topological string theory is to apply a topological twist, which is an operation invented by Witten that alters the representations of various fields under the 2D Lorentz group. The topological twist effectively decouples the dynamical degrees of freedom; the resulting theory is wholly topological. Further, the model needs to be integrated over the moduli space of the worldsheet to couple the fields to gravity. Anomalies are best avoided, and the most interesting observables result when the target space is a Calabi-Yau threefold. The topological string amplitude $\mathcal{F}^{(g)}$, which is formally defined as the path integral for a worldsheet of genus $g$, can be summed into a generating function known as the topological-string free energy $\mathcal{F} = \sum_{g=0}^{\infty} \mathcal{F}^{(g)} \lambda^{2g-2}$. Here, the parameter $\lambda$ serves as the string-coupling constant. Sometimes, the topological string partition function $Z=\exp (\mathcal{F})$ is more convenient to compute than the free energy.

The twisting operation mixes the 2D Lorentz group with a $U(1)$ global symmetry of the worldsheet theory, which can be  either vector symmetry or axial symmetry. The two choices are known as A-twisting and B-twisting respectively. The resulting  topological string theories are known as A-and B-models. The A-model counts holomorphic curves in the Calabi-Yau threefolds, and its rigorous mathematical formulation is known as the Gromov-Witten theory \cite{Ruan:1990kt, Kontsevich:1994na}. While the Gromov-Witten invariants are rational numbers, the expansion of the generating function in terms of GopakumarÐVafa invariants is more appealing \cite{Gopakumar:1998ii}. These invariants are integers that count the BPS particles in the five dimensional theory of M-theory on Calabi-Yau threefolds. In contrast, the B-model probes the Hodge deformation of Calabi-Yau threefolds. The partition functions for the A-model and B-model depend  on the Kahler and complex-structure moduli of the Calabi-Yau threefold, respectively, and they compute certain holomorphic terms in the four-dimensional $\mathcal{N}=2$ supergravity effective action of type II A/B superstring theories compactified on Calabi-Yau threefolds.

Mirror symmetry states that an A-model on a Calabi-Yau threefold is equivalent to a B-model on a mirror Calabi-Yau threefold that exchanges the Hodge numbers. Then, the count of genus-zero curves, which are also known as rational curves, on the A-model side can be transformed to the simpler problem of Hodge deformation on the B-model side, which is described by the PicardÐFuchs linear differential equation. These predictions motivated developments of the mathematical method of localization in moduli space to calculate Gromov-Witten invariants and have been proven rigorously by mathematicians \cite{Givental, Lian:1999rn}.

Over the years, many techniques have been developed to compute topological string partition functions beyond the genus-zero sector. For a class of non-compact toric Calabi-Yau threefolds, the geometries can be described by trivalent toric diagrams, and the partition function can be determined completely through the physical method of pasting together the topological vertices \cite{Aganagic:2003db}. Furthermore, these physics-based results have been proven rigorously with mathematical methods \cite{Li:2004uf}.

The situation for computing higher genus topological string amplitudes is far more difficult for compact Calabi-Yau threefolds. In this case, the most effective method is the holomorphic anomaly equation given by Bershadsky, Cecotti, Ooguri, and Vafa (BCOV)\cite{Bershadsky:1993cx}. This equation can be derived using the path integral formalism, and it recursively relates the anti-holomorphic derivatives of high-genus topological string amplitudes in terms of lower-genus amplitudes owing to the degenerations of the string worldsheet. Furthermore, the topological string amplitudes can have regular poles on special points where BPS particles become massless. It is evident that the BCOV equation does not fix any meromorphic part of the topological string amplitudes. It is well known in complex analysis that while on a non-compact complex manifold such as $\mathbb{C}$, there are infinitely many linearly independent holomorphic functions, on a compact complex manifold such as $\mathbb{CP}^1$, a holomorphic function must be a constant function.   
 In the B-model, the topological string amplitudes depend on the complex structure parameters of the Calabi-Yau manifolds, which live in a compact moduli space by including points at infinity. If only regular poles of a certain degree appear at the special singular points of the complex structure moduli space, the high-genus topological string amplitudes are determined recursively up to a meromorphic function of finite unknown constants, often referred to as holomorphic ambiguity. These unknown constants need to be fixed by other means. For example, although no systematic method is available to calculate the GopakumarÐVafa invariants, geometric methods for their computation can be developed in special cases. Using these geometric results, the topological string amplitudes for the quintic, a well-known compact Calabi-Yau threefold, were calculated up to genus 3 by Katz \cite{Katz:1999xq}.

We reported significant progress on this problem in 2006 \cite{Huang:2006hq}. Firstly, we use a formalism that takes topological string amplitudes as polynomials, drawn from Yamaguchi and Yau \cite{Yamaguchi:2004bt}. This polynomial formulation greatly improves the efficiency of integrating the holomorphic anomaly equation. Furthermore, we proposed novel boundary conditions for higher-genus topological string amplitudes by considering the effective actions of integrating out the nearly massless particles around singular points of the moduli space. These conditions are particularly useful near the conifold point of the Calabi-Yau moduli space and are sometimes referred to as the gap condition, owing to the gap-like pattern of the series expansion in terms of mirror coordinates. For non-compact Calabi-Yau threefolds, these boundary conditions are usually sufficient to fix the topological string amplitudes to all genera. However, in the compact case, the conditions are still insufficient at some high genera. For example, for the quintic threefold, we predict that our boundary conditions can fix the topological string amplitudes up to genus 51; in practice, we were able to compute the theory up to approximately genus 20 with everyday computers. Our results were tested with many geometric calculations of the GopakumarÐVafa invariants.

On the mathematical side, only genus one formulas have been proven so far for the compact case  \cite{Zinger}. Mathematicians have been making tremendous heroic efforts to push beyond to higher genera.

The rest of this review article consists of two main parts. In Section \ref{section2} we review more recent progress on computing topological string amplitudes on a class of elliptic Calabi-Yau manifolds \cite{Huang:2015sta}. Motivated by earlier work, due to the elliptic fibration structure, we propose that an ansatz of the topological string-partition function can be written in terms of Jacobi forms. In concert with our previously reported techniques \cite{Huang:2006hq}, we can, in principle, compute the topological string amplitudes to much higher genera for a compact example of elliptic fibration over $\mathbb{P}^2$; however, the calculation will not cover all genera and Kahler classes. Additionally, the new techniques can be applied to compute elliptic genera of 6D superconformal field theories related to non-compact elliptic Calabi-Yau manifolds and often provide complete solutions in these cases.

This review article consists of two main parts. In Section \ref{section2} we review the more recent progress on computing topological string amplitudes on a class of elliptic Calabi-Yau manifolds \cite{Huang:2015sta}. Motivated by earlier works, due to the elliptic fibration structure, we propose the topological string partition function can be written by an ansatz  in terms of Jacobi forms. Combined with the old techniques in  \cite{Huang:2006hq}, we are able to compute the topological string amplitudes in principle to much higher genus for a compact example of elliptic fibration over $\mathbb{P}^2$, though still not completely to all genera and all Kahler classes. The new techniques can be also applied to compute elliptic genus of 6d superconformal field theory related to non-compact elliptic Calabi-Yau manifolds, and often provide complete solutions in these cases. 

In Section \ref{section3} we discuss another aspect of the potential applications of topological string theory. In quantum mechanics, exactly solvable systems are rather rare. Recently,  topological strings have been found to provide exact quantization conditions, including non-perturbative contributions, for a class of quantum-mechanical systems derived from the mirror curves of toric Calabi-Yau manifolds. These surprising connections shed new light on studies of integrable models in mathematical physics.

\section{Elliptic Calabi-Yau Manifolds and Jacobi Forms} \label{section2}

Elliptic fibration appears naturally in F-theory \cite{Vafa:1996xn}, which is essentially type IIB string theory with the axion-dilaton field varying over a compact internal space. Due to the $SL(2,\mathbb{Z})$ duality, the axion-dilaton field can be identified with the modulus of a torus and plays the role of elliptic fiber parameter. To construct phenomenologically interesting models that resemble our 4d world, one would consider F-theory compactified on Calabi-Yau four-folds. 

In the article we consider Calabi-Yau threefolds, and the F-theory compactification gives rise to a 6d theory. In the case where a compact cycle in the base of the elliptic Calabi-Yau threefold is contractible, we can take a local limit and decouple gravity. The resulting 6d  superconformal field theories (SCFT) attracted some renaissance of research interests in recent years, and have been systematically classified in the context of F-theory compactification. See e.g. \cite{Heckman:2018jxk}  for a recent review. In dimension higher than 4, the gauge coupling constant has positive mass dimension, so gauge theory is not renormalizable in the UV and free in the IR. It is difficult to write down a Lagrangian for an interacting conformal field theory in these high dimensions. These 6d SCFTs came as a surprise when they were first constructed in string theory in middle 1990's. They have tensionless string degrees of freedom and it is generally believed that they can not be described by a local Lagrangian as in conventional quantum field theory. 

Two especially interesting examples of 6d SCFTs can be also constructed in M-theory. It is a well known string duality that M-theory compactified on a line interval with two M9-branes at each end points is dual to the weakly coupled $E_8\times E_8$ heterotic string theory \cite{Horava:1995qa} when the size of the line interval goes to zero. Each ``end of the world" M9-branes provides a $E_8$ gauge symmetry, and M2-branes stretching between the two M9-branes becomes heterotic strings. Alternatively, we have also M2-branes stretching between two M5-branes, or M2-branes stretching between one M5-brane and one M9-brane. They become strings on the worldsheet of  the M5-brane, are known as M-string and E-string respectively.  

One can define elliptic genus for these 6d SCFTs. Consider the 2d worldsheet theory of the tensionless strings, the elliptic genus is defined similarly as string partition function, by trace over the Hilbert space, with the extra insertions of $(-1)^F$ operator that change sign for fermionic sector, and some extra ``fugacities" from symmetries of the theory. In path integral formalism, one integrates the field configurations over a torus, with the $(-1)^F$ operator changing boundary conditions for fermionic fields. 

Through a chain of duality, the refined topological string partition functions on local elliptic Calabi-Yau threefolds is equivalent to the elliptic genus of the corresponding 6d SCFTs from F-theory compactification. The elliptic genus of M-strings are computed by refined topological vertex method \cite{Iqbal:2007ii, Haghighat:2013gba}. On the other hand, the elliptic genus of E-string corresponds to the topological string partition function of a local half K3 Calabi-Yau threefold, which is a local limit of the elliptic fibration over $\mathbb{F}_1$ Hirzebruch surface. Inspired also by refined topological vertex formalism, the elliptic genus of E-string is written in terms of Jacobi form ansatz in \cite{Haghighat:2014pva}. Utilizing previous results \cite{Huang:2013yta}, the ansatz for elliptic genus of two E-strings can be completely fixed, and the calculations were subsequently pushed to three E-strings \cite{Cai:2014vka}. A more powerful method is to construct the worldsheet theory for E-strings, and use supersymmetric localization to calculate the path integral of the elliptic genus exactly \cite{Kim:2014dza}. This method in principle can calculate the elliptic genus of any number of E-strings. However, for general  6d SCFTs it is quite difficult to construct the worldsheet theory, and only sporadic examples are known to date.

\subsection{A compact model: elliptic fibration over $\mathbb{P}^2$} 

Motivated by these developments, we apply the Jacobi form ansatz technique to compact Calabi-Yau threefolds \cite{Huang:2015sta}. One of the simplest examples is the degree 18 hypersurface in weighted projective space, usually denoted as  $X_{18} (1,1,1,6,9)$, which can be also realized as an elliptic fibration over $\mathbb{P}^2$. The geometry has two Kahler parameters, one from the base and another from the fiber, and was studied in the early days of mirror symmetry \cite{Candelas:1993dm}. 

The Calabi-Yau geometry and its mirror are described by a dual pair of reflexive lattice polytopes $(\Delta, \Delta^*)$, due to Batyrev \cite{Batyrev}. In general, for compact Calabi-Yau $n$-folds we consider polytopes in $\mathbb{R}^{n+1}$. For simplicity we can define an inner product in $\mathbb{R}^{n+1}$ to be the usual dot product, then the dual polytope is defined also in $\mathbb{R}^{n+1}$ as $\Delta^{*} =\{ y\in \mathbb{R}^{n+1} | x\cdot y \geq -1, \forall x\in \Delta \}$. The reflexive polytopes $(\Delta, \Delta^*)$ both have vertices at integer lattice points, and only one internal integer lattice point, usually chosen to be the origin. For $n=1$ such reflexive polytope pairs were completely classified by 16 polygons in $\mathbb{R}^{2}$, while for higher dimension the classifications are much more complicated. Mirror pairs of Calabi-Yau threefolds with elliptic fibration structure can be constructed by two pairs of two dimensional reflexive polygons, with each pairs serving as the base and the elliptic fiber, and fitting together into a pair of reflexive polytopes in $\mathbb{R}^{4}$. 

We denote the two complex structure parameters of the mirror by $z_1, z_2$. Following the standard procedure of mirror symmetry one can write the Picard-Fuchs differential equations and solve for the periods and the mirror maps. Here the two mirror maps $t_i\sim \log(z_i)$ are identified with the Kahler parameters of the base and the fiber in $X_{18} (1,1,1,6,9)$.  We can also determine the three point functions and the leading term of the topological string free energy, known as the prepotential $\mathcal{F}^{(0)}$ from the period vectors. 

Early works discovered an involution symmetry in the complex structure moduli space of the mirror 
\begin{eqnarray}
I: ~~ (z_1, z_2) \rightarrow (\frac{1}{432} -z_1, -\frac{(432z_1)^3z_2}{(1-432z_1)^3} ) .
\end{eqnarray} 
The Picard-Fuchs equations are invariant up to some trivial factors under the involution, and up to some simple factors,  the involution symmetry exchanges the two following conifold discriminants of the Picard-Fuchs equations 
\begin{eqnarray}
\Delta_1 &=& (1-432z_1)^3 -27 z_2 (432z_1)^3, \nonumber \\  \Delta_2 &=& 1+27z_2. 
\end{eqnarray}

It turns out the involution symmetry is useful for constraining the higher genus topological string amplitudes. The involution symmetry acts on the holomorphic 3-form by $I: \Omega \rightarrow i\Omega$, and $\Omega$ defines the vacuum line bundle $\mathcal{L}$. The higher genus topological amplitude $\mathcal{F}{(g)}$ transforms as a section of $\mathcal{L}^{2g-2}$, so it must transform as $I: \mathcal{F}{(g)}\rightarrow (-1)^{g-1} \mathcal{F}^{(g)}$. 

The higher genus topological string amplitudes for one-parameter Calabi-Yau models such as the well-known quintic manifold can be written as polynomials of certain generators with real coefficients \cite{Yamaguchi:2004bt}. For multi-parameter models, it is more convenient to write the amplitudes as polynomials of BCOV propagators \cite{AL} with a shift, denoted as $S^{ij}, S^i, S$ where the index labels the complex structure parameters, and the coefficients of the polynomials are now rational functions of complex structure parameters, with poles only at the discriminants. The BCOV propagators are not holomorphic and satisfy a set of holomorphic anomaly equations. Usually these equations do not completely determine the BCOV propagators and one has the freedom to choose some holomorphic ambiguities consistent with the equations. For the $X_{18} (1,1,1,6,9)$ model here, this was solved in previous work \cite{Alim:2012}. We use the gauge choice in \cite{Alim:2012} and determine how the  BCOV propagators transform under the involution symmetry. It turns out that the BCOV propagators do not transform as a tensor, but with some shift of rational functions. As a polynomial of BCOV propagators, we then deduce the transformation of $\mathcal{F}^{(g)}$, which constrain the holomorphic ambiguity for $\mathcal{F}^{(g)}$, and reduced the number of unknown coefficients by about one quarter.

It also turns out the involution symmetry is intimately related to a modularity structure in earlier literatures. In the study of the $\mathcal{N}=4$ topological Yang-Mills theory, it was realized the partition function can be written as quasi-modular forms, which are homogeneous polynomials of Eisenstein series $E_2, E_4, E_6$ with modular weights $2,4,6$ respectively  \cite{Minahan:1997}. They are call quasi-modular since it is well known that $E_2$ is not modular, but transforms with a shift under the S-transformation $\tau\rightarrow -\frac{1}{\tau}$.  An alternative choice is to shift the $E_2$ by an anholomorphic piece so that the shifted $E_2$ becomes modular. The ring of quasi-modular forms is closed under the derivative of modular parameter $\tau$, due to the Ramanujan formulas. It was observed that the derivative of the partition function as a polynomial with respect to $E_2$, can be written in terms of lower order terms. This is called a ``modular anomaly equation", since a true modular form would be simply polynomials of $E_4, E_6$, and would vanish under such derivative of $E_2$. The theory of quasi-modular forms is described in the book \cite{Zagier2008}, and also appeared in the partition function of 2d Yang-Mills theory \cite{Dijkgraaf, Kaneko}, although in which case there appears to be no simple modular anomaly equation.

The $\mathcal{N}=4$ topological Yang-Mills is described by the genus zero sector of the E-string theory. Some generalizations of the modular anomaly equations are subsequently proposed, to including $E_8$ mass parameters \cite{MNVW}, to higher genus E-strings \cite{HST}, to refined theory \cite{Huang:2013yta}, to compact elliptic Calabi-Yau models \cite{Klemm:2012}. In general elliptic Calabi-Yau models, this quasi-modularity structure comes from the elliptic fiber. For our main model $X_{18} (1,1,1,6,9)$, we find that the involution symmetry constrains on the holomorphic ambiguity is essentially equivalent to the elliptic fiber modularity. 

These modular anomaly equations are precursors for the use of weak Jacobi forms. Jacobi forms are holomorphic functions of two variables, one modular parameter $\tau$ on the upper half plane as in the case of modular forms, and another elliptic parameter $z$ which is quasi-periodic on the complex plane on lattice generated by $1,\tau$. Two integers appearing in the transformation rules are known as the modular weight and index. The modular forms $E_4, E_6$ can be thought of as a special Jacobi forms of weights $4,6$ and index $0$. One can impose some choices of  the vanishing conditions on Fourier expansion in $e^{2\pi i \tau}, e^{2\pi i z}$, and called such forms (holomorphic) Jacobi forms, weak Jacobi forms, etc. We should clarify that the weak Jacobi forms are also holomorphic functions, but just have a weaker condition on the Fourier coefficients that the (holomorphic) Jacobi forms. 

The theory of Jacobi forms was systematically studies in the book \cite{EichlerZagier}. Like modular forms, the space of holomorphic or weak Jacobi forms of given weight $k$ and index $m$, denoted $J_{k,m}$ or $\tilde{J}_{k,m}$, is a finite dimensional linear space over $\mathbb{C}$. We can consider the graded rings which are formal sums of Jacobi forms of different weights and indices. For the (holomorphic) Jacobi forms, the graded ring of fixed index $J_{*,m}$ is a free module  of rank $m+1$ over the ring of $SL(2,\mathbb{Z})$ modular forms $M_{*}$. Namely, there are $m+1$ generators, so that an index $m$ (holomorphic) Jacobi form can be uniquely written as a linear combination of the $m+1$ generators with modular form coefficients. Furthermore, the $m+1$ generators can be constructed as polynomials of two special generators with meromorphic modular form coefficients with only poles from the discriminant $E_4^3-E_6^2$. However, considering varying also the index, then the bigraded ring $J_{*,*}$ is not finitely generated. There is not a finite number of (holomorphic) Jacobi forms that generate all (holomorphic) Jacobi forms as their polynomials. We can instead work in a larger space of weak Jacobi forms, where the situation is better that the bigraded ring $\tilde{J}_{*,*}$ is  finitely generated. In fact, it is freely generated by polynomials of 4 generators $E_4, E_6, \phi_{-2,1}, \phi_{0,1}$, where $\phi_{-2,1}, \phi_{0,1}$ can be constructed from Jacobi theta functions and the subscript denotes their modular weight and index. For topological string amplitudes we will mainly use this weak version. Even if sometimes the amplitudes are actually (holomorphic) Jacobi forms, it is still more convenient to work with the weak version since it is easier to write ansatz with any index. 

In the small $z$ expansion, it is well known in the theory of Jacobi forms that the coefficients are quasi-modular forms, and satisfy a modular anomaly equation according to its index. So if we identify the elliptic parameter $z$ with the topological string coupling constant, it seems to automatically imply the modular anomaly equations of the early literatures. This works out for the non-compact model of E-string theory \cite{Haghighat:2014pva}. One especially powerful feature is that the weak Jacobi forms fix the modular ambiguity. In order to solve topological string amplitudes by modular anomaly equation, one has to fix a purely modular piece which can be added without affecting the modular anomaly equation. This modular part is called modular ambiguity. In simple low order amplitudes, one can sometimes fix the modular ambiguity by conditions from vanishing Gopakumar-Vafa invariants. However, this becomes increasingly difficult for higher order amplitudes. So the use of weak Jacobi forms enables one to fix these modular ambiguities for all genus amplitudes, for a given base Kahler class. 

We propose that the topological partition function of elliptic fibration over $\mathbb{P}^2$ can be written as 
\be 
Z=Z_0( (\tau,z) (1+ \sum_{d_B =1}^{\infty} Z_{d_B} (\tau,z) Q^{d_B} )\ ,
\label{expansion} 
\ee
where $z$ is identified with the string coupling constant, $Q$ is the exponential of base Kahler parameter of $\mathbb{P}^2$ with a shift, and $\tau$ is the fiber Kahler parameter. The prefactor $Z_0(\tau,z)$ is independent of the base and can be determined in terms of modular forms. The ansatz for $Z_{d_B} (\tau,z)$ is not holomorphic, but only meromorphic, with poles from the structure of Gopakumar-Vafa expansion of topological string amplitudes. We use a simple universal denominator, motivated by previous calculations on E-string theory, to capture all the possible poles. The numerator has no pole and is a weak Jacobi form of certain modular weight and index. We have the ansatz 
\be 
Z_{d_B}(\tau,z)=\frac{\varphi_{d_B}(\tau,z)}{\eta^{36 d_B}(\tau)\prod_{k=1}^{d_B} \varphi_{-2,1}(\tau,k z)}. 
\label{ansatz}
\ee
The ansatz has formally zero modular weight, so the weight of the numerator can be easily deduced from the denominator. The index can be also determined from the modular anomaly equation. So the numerator can be written as a polynomials of 4 generators $E_4, E_6, \phi_{-2,1}, \phi_{0,1}$ with a finite number of coefficients. 

A large number of the coefficients in the ansatz of the numerator $\varphi_{d_B}(\tau,z)$ can be fixed by vanishing constrains of Gopakumar-Vafa invariants. As we mentioned, the Gopakumar-Vafa invariants count BPS particles from M2-branes wrapping two-cycles in Calabi-Yau compactification. The spin of the BPS particles corresponds to genus in topological string amplitudes and has an upper bound for a given Kahler class. For base degree $d_B$, the contributions to Gopakumar-Vafa invariants come from not only $Z_{d_B}$ but also lower base degree, known as multi-cover contributions. The ansatz of $Z_{d_B}$ for $d_B>1$ contributes to Gopakumar-Vafa invariants of arbitrary high spin (genus) with a Kahler class of fixed base and fiber degrees. For our conjecture to be valid, there should be at least one particular ansatz that these non-vanishing Gopakumar-Vafa invariants can be cancelled from lower degree multi-cover contributions. We assume this is indeed the case, and ask what can be add to this particular ansatz such that there is no additional contributions to Gopakumar-Vafa invariants of arbitrary high spin. We call an element in the remaining degrees of freedom, a linear space over $\mathbb{C}$, the \textit{restricted ansatz}. It turns out that the factor $\varphi_{-2,1}(\tau,k z)$ in the denominator with $k>1$ always contributes to Gopakumar-Vafa invariants of arbitrary high spin, while the first factor $\varphi_{-2,1}(\tau, z)$ does not. So the restricted ansatz is simply of sub-family of the ansatz (\ref{ansatz}) with only $\eta^{36 d_B}\varphi_{-2,1}(\tau, z)$ in the denominator. As the overall modular weight and index remain the same, this vanishing geometric constrains significantly reduce the weight and index of the numerator, and its number of coefficients to fix.    

The ansatz is similar to formulas from A-model topological string theory, like the topological vertex, in that it computes topological amplitudes for a fixed Kahler class, but to all genera. Here we compute recursively with increasing base degree $d_B$, relying on the lower degree results for imposing the vanishing Gopakumar-Vafa constrains. We combine this approach with the B-model holomorphic anomaly approach, where we compute recursively with increasing genus, but the results for $\mathcal{F}^{(g)}$ is valid at all points in the complex structure moduli space of the mirror, and we can obtain all Gopakumar-Vafa invariants of genus $g$ for all Kahler class.  

After some careful analysis, using the involution symmetry and the boundary conditions at the conifold point,  we find that the exact formula at base degree $d_B$ can provide sufficient boundary data to fix the B-model formula up to genus $9(d_B+1)$. On the other hand, in order to fix the exact formula at base degree $d_B$, we need topological free energy of genus no less than $\frac{d_B(d_B-3) }{2}+1$.  Thus, as long as $9(d_B+1) \geq \frac{(d_B+1)(d_B-2) }{2}+1$, we can repeat this procedure to fix the exact formula with increasing base degrees. If no other obstructions appear, in this way we can in principle determine the exact formula up to base degree $d_B=20$ (for all genera and fiber degrees), and the topological string free energy up to genus 189 (for all base and fiber degrees). In practice we compute up to $d_B=5$ and genus $g=8$. Although we think these are significant progress over previous works, we still fall short of completely solving topological string theory on a generic compact Calabi-Yau threefold for all Kahler class and all genera. This remains an interesting guiding problem for future works. 

\subsection{Elliptic genus of 6d SCFTs} 

As we mentioned that 6d SCFTs can be constructed from F-theory on non-compact elliptic Calabi-Yau threefolds. It is conjectured by Vafa and collaborators that such constructions provide a complete classifications of $(1,0)$ SCFTs in 6d, though it is difficult to prove the claim from bottom-up field theoretical arguments. There have been tremendous research works on classifications of such constructions, and the compactifications to lower dimensions. Since these theories are still mysterious from purely conventional field theory point of view, due to their non-Lagrangian nature, it is worthwhile to study calculable physical quantities in these theories. 

Here we focus on some simplest cases, known as the minimal 6d SCFTs \cite{Haghighat:2014vxa}. They have rank one in the tensor branch and realized as F-theory compactified on elliptic fibration over a non-compact base $\mathcal{O} (-n)\rightarrow \mathbb{P}^1$, where $n=1,2,\cdots 8, 12$.  These 9 theories serve as the atomic building blocks, and other $(1,0)$ SCFTs can be constructed by pasting together these minimal theories according to certain rules. 

Elliptic fibration becomes singular when the elliptic fiber becomes degenerate at some points in the base. As an elliptic curve, we describe the fiber in terms of Weierstrass form and the degenerate discriminant loci are generically a divisor hypersurface in the base. There are now standard mathematical procedures to resolve these singularities, classified by Kodaira in terms of Lie algebra. On the other hand, in physics language, we introduce D7 branes to wrapping on the singular locus in the base, and extend over the 6 uncompactified dimensions, giving rise to gauge symmetry in the SCFTs.  Unlike gauge symmetry from open strings ending on freely moving D-branes, which can be Higgsed and broken by moving D-branes apart. Here the gauge symmetry from elliptic fibration singularity comes from Lie algebra in Kodaira classification and is not Higgsable. In this sense these theories are truly minimal and can not be reduced further by Higgs mechanism. 

There are tensionless strings in these 6d SCFTs from D3 branes wrapping the collapsed cycle in the base of the elliptic Calabi-Yau threefold. In principle there is a worldsheet theory with supersymmetry on one side so we can compute elliptic genus of the tensionless strings. However it is difficult to directly deal with tensionless strings. Usually one constructs a worldsheet theory with intersecting D-branes, whose Lagrangian can be explicitly written, and which flows to the theory of tensionless strings in the infrared. The elliptic genus as protected by supersymmetry is invariant under the renormalization group flow so we can compute the elliptic genus with the simpler worldsheet theory. 

Let us discuss some basic properties of elliptic genus. Mathematically, elliptic genus can be defined for a manifold as integral in terms of Chern class. It has two parameters and transforms similarly as Jacobi forms for Calabi-Yau manifolds. Here we are interested in the physical definition, due to Witten \cite{Witten:1986bf}. Recall first the simpler case of  supersymmetric quantum mechanics, where the Hilbert space can be separated into the bosonic part and the fermionic part. The Witten index is defined as $\Tr(-1)^Fe^{-\beta H} $ as the usual partition function with an additional fermionic operator $(-1)^F$  that changes the sign of contributions from the fermionic sector. Supersymmetric quantum mechanics has non-negative energy spectrum. For strictly positive energy eigenvalue, it can be shown there is an isomorphism between bosonic sector and the fermionic sector. So the Witten index receives contributions  only from the ground state $\Tr(-1)^Fe^{-\beta H} = \Tr(-1)^F$, and is independent of the temperature parameter $\beta$. This can be also derived in the path integral formalism by integrating with the Euclidean action over the thermal circle of length $\beta$. The effect of the $(-1)^F$ operator is to change the boundary condition for fermions from anti-periodic to periodic. For a manifold $M$, we can construct a supersymmetric quantum mechanics with $M$ as the target space, then the Witten index is exactly the Euler character of the manifold $M$. 

For elliptic genus we consider two-dimensional theory with extended supersymmetry on right moving sector but not left moving sector. The result agrees with the mathematical definition of the elliptic genus of a manifold $M$ for the non-linear sigma model with $M$ as the target space, but we can consider more general theories. As in usual canonical quantization of string theory, we compactify the spatial direction of the worldsheet so we have a quantum mechanics on the time direction and a Hilbert space generated by string excitation modes. We can impose periodic or anti-periodic boundary condition for fermions which only appears in the right moving sector here, giving rise to Ramond or Neveu-Schwarz(NS) sectors. Here we consider the elliptic genus of Ramond sector though it is also possible to study the NS sector. Now we can define the flavored Ramond elliptic genus similarly as Witten index in terms of trace over Hilbert space 
\begin{eqnarray} \label{ellipticgenus}
&& Z_\beta(\tau, \epsilon_1,\epsilon_2,{\bf m}) \\ \nonumber &\equiv& \text{Tr}_R \, (-1)^F q^{H_L} \bar q^{H_R} u^{2J_-} v^{ 2(J_+ + J_r) }  \,{\bf e}[{\bf m}] . 
\end{eqnarray}
We give some explanations of the notations here. The trace is over Ramond sector denoted by the subscript R. The elliptic genus is a function of parameters $q=e^{e\pi i \tau}, u = e^{2\pi i \epsilon_{-}}, v=e^{2\pi i \epsilon_{+}}$, and mass parameters. The $H_L$ and $H_R$ are the left moving and right-moving Hamiltonians, $J_{+}, J_{-}$ are the generators of the decomposition of spatial rotation group $SO(4)\simeq SU(2)\times SU(2)$ of the tensionless strings in 6d, and $J_r$ is the R-symmetry generator of the supersymmetry algebra in the right moving sector. The 6d SCFT may have gauge or global symmetries that both become global symmetry in the worldsheet theory, and we can insert the factor  ${\bf e}[{\bf m}]$ which is the exponential of the generators of these symmetries with some mass parameters. Furthermore, the strings have charge valued on the lattice of the contractible compact cycles, which is denoted by $\beta$. For the minimal rank one theories, the charge $\beta$ is simply a natural number representing the number of strings. 

Sometimes an explicit worldsheet theory, usually a quiver type gauge theory from intersecting branes, can be constructed which flow to the strings of 6d SCFT in the infrared. One may try to determine the string spectrum and compute the trace in (\ref{ellipticgenus}). However this seems difficult and we don't know of research in this direction. Instead it is better to compute the elliptic genus by path integral formalism. Now we also compactify the time direction on a thermal circle and as before the $(-1)^F$ operator change fermion boundary condition from anti-periodic to periodic. So we compute path integral of the remaining fugacity operators on a torus of complex modulus $\tau$ with periodic boundary conditions for fermions in both compact directions. The computation can be done by techniques of supersymmetry localization, which reduce the path integral to around the loci where the supersymmetry transformation of fermions vanishes. One expands around such loci and compute the one-loop determinant from only quadratic terms, which gives the exact results due to arguments from supersymmetric deformation invariance. Usually the path integral localizes to some residue calculations, and the results can be written in terms of Jacobi theta function \cite{Kim:2014dza, Haghighat:2014vxa, Kim:2016foj} 

The generating function summing over the elliptic genus of all charges $\beta$ is essentially the partition function of 6d SCFTs on $\mathbb{R}^4_{\Omega} \times T^2$. The generating parameter for string charge $\beta$ is identified with the vev of scalar in the tensor branch, when one moves away from the tensionless point. Here $\mathbb{R}^4_{\Omega}$  is the Omega deformation of $\mathbb{R}^4$, twisting the two rotations by two parameters $\epsilon_{\pm} =\frac{1}{2} (\epsilon_1\pm \epsilon_2)$. The Omega background was first introduced by Nekrasov \cite{Nekrasov:2002qd}, to regularize the integrals over moduli space of instantons in 4d $\mathcal{N}=2$ Seiberg-Witten theory. The the leading term of Nekrasov partition function is the prepotential in Seiberg-Witten theory, and we also have higher order terms from two expansion parameters $\epsilon_{1,2}$, which are interpreted as gravitational couplings of gauge theory and compute the effective action of $R^2$  coupling to graviphoton field strength. Motivated by the gauge theory calculations, one defines the \textit{refined} topological string theory with two coupling constants $\epsilon_{1,2}$.  Here we can think in terms of the spacetime effective action. Conventional topological string free energy of genus $g\geq 1$ on Calabi-Yau threefolds compute the effective actions term $R^2 F^{2g-2}$ in type IIA compactification to 4d, where $F$ is the self-dual part of graviphoton field strength \cite{Antoniadis:1993ze, Bershadsky:1993cx}. Now for refined topological string theory, we allow for both self-dual and anti-self-dual  graviphoton field strength and get a two-parameter expansion \cite{Iqbal:2007ii}. This seems possible at least for non-compact Calabi-Yau threefolds where one always has $U(1)$ isometry in the geometry. The refined Gopakumar-Vafa invariants $n^{\beta}_{j_L,j_R}$ count 5d BPS particles with both $SU(2)$ spins from the decomposition of rotation group $SO(4)$, instead of an alternating sum over the right spin in the unrefined case. As a result, the unrefined Gopakumar-Vafa invariants may be negative integers but the refined Gopakumar-Vafa invariants are always non-negative. Although naively it seems difficult to understand a two parameter expansion from worldsheet genus expansion of string theory, such a realization is discussed in \cite{Antoniadis:2013epe}.

We mentioned earlier that the elliptic genus is now identified with refined topological string partitions on the corresponding elliptic Calabi-Yau threefolds with the refined string couplings $\epsilon_{1,2}$. The string charge $\beta$ is identified with the Kahler class in the base, while the complex structure parameter of the torus $\tau$ is identified with the Kahler parameter of the elliptic fiber. Furthermore, the mass parameters from global or gauge symmetry are identified with Kahler parameters for certain non-compact cycles from resolution of the elliptic fiber singularity. 

We can also apply the Jacobi form ansatz method to the minimal theories in \cite{Haghighat:2014vxa}. For $n=1, 2$ cases, the theories are simply the E-strings, M-strings theories mentioned earlier that can be also constructed in M-theory. For these two cases there is no non-Higgsable gauge symmetry, and there is a global $E_8$ symmetry in the E-string theory. The $n=2$ case of  M-string theory actually has enhanced $(2,0)$ supersymmetry. For the other cases there are non-Higgsable gauge symmetries from singular elliptic fiber. In the compact case we only consider the unrefined theory, but here we can make an ansatz for the refined theory \cite{DelZotto:2016pvm, Gu:2017ccq, DelZotto:2017mee}. 

The ansatz for theories without gauge symmetry is 
\begin{eqnarray} \label{EMstrings}
&& Z_{\beta}  = \left(\frac{\sqrt{q}}{\eta(\tau)^{12}}\right)^{-\beta \cdot K_B}
\\ \nonumber &\times &
\frac{\phi_{k,\np,\nm,\beta}(\tau,\boldsymbol{m},\epsilon_+,\epsilon_-)}{ \prod_{i=1}^r \prod_{s=1}^{\beta_i} \left[ \phi_{-1,\frac{1}{2}}(\tau, s \epsilon_1) \phi_{-1,\frac{1}{2}}(\tau, s\epsilon_2) \right]} \,,
\end{eqnarray}
Here are some explanations of the notations. The prefactor is due to a shift of Kahler class in the base. $\beta$ denotes the Kahler class in the base. For E-strings and M-strings, there is only one class in the base so $r=1$ and $\beta=n$ is the number of strings or the degree of Kahler class. The denominator accounts for all possible poles from the refined topological string theory. The numerator $\phi_{k,\np,\nm,\beta}$, to be fixed, is a polynomial of weak Jacobi forms of appropriate weight and indices in $\epsilon_{\pm} = \frac{1}{2} (\epsilon_1\pm \epsilon_2)$, and the $E_8$ Weyl invariant weak Jacobi forms (E-string) or weak Jacobi forms of a mass (M-string). We assume here that the weak Jacobi forms with multiple sets of elliptic parameters can be always written in terms of weak Jacobi forms of individual set of parameters with each index added up independently, so the numerator can be written as an ansatz of known generators with finite number of unknown coefficients. 

The conventional Jacobi forms with one elliptic parameter can be generalized to include a simple Lie algebra. The number of elliptic parameters is the rank of the simple Lie algebra, parametrizing a vector in the space spanned by the simple roots. Besides similar transformation properties, we also require the Jacobi forms to be invariant under Weyl group transformations of the root system. This is known as Weyl invariant Jacobi forms. The conventional Jacobi forms can be thought of as a special case of $SU(2)$ Jacobi forms which is the only rank one case. Like in the previous case, here we also consider the weak version of Jacobi forms. For all simple Lie algebra except the $E_8$ case, the ring structure is quite similar to conventional case. it was shown in \cite{Wirthmuller} that there are a finite set of generators whose polynomials freely generate the ring of Weyl invariant weak Jacobi forms. Explicit generators for some cases were constructed in \cite{Bertola}. 

On the other hand, the $E_8$ case is quite exceptional. Sakai constructed 9 holomorphic Jacobi forms $A_n, B_n$ of weight $4,6$ and positive indices $n$. For a fixed index, it is proven that like holomorphic Jacobi forms, the weak Jacobi forms form a free module over the ring of modular forms $M_{*}$  \cite{Wang:2018fil}. The rank of the module, namely the number of independent generators of the fixed index, is given the number of monomials of Sakai generators of the same index. Furthermore, the basis of generators can be always constructed by polynomials of Sakai generators with meromorphic modular form coefficients. Some basis for both holomorphic and weak Jacobi forms in the cases of indices less than 5 are constructed in \cite{Wang:2018fil}. Here an interesting feature appears at index 5. In the calculations on E-strings \cite{Huang:2013yta, DelZotto:2017mee}, we encounter a special polynomial of index 5 and weight 16
\begin{eqnarray} \label{polynomialB.35} \nonumber 
P && =  864 A_1^3 A_2+21 E_6^2 A_5 -770 E_6 A_3 B_2 \\ && +3825 A_1 B_2^2  -840 E_6 A_2 B_3+60 E_6 A_1 B_4.  \nonumber 
\end{eqnarray}
We check numerically this polynomial vanishes at the zero points $\tau=\pm \frac{1}{2}+\frac{\sqrt{3}}{2}i$ of $E_4$ for general $E_8$ mass parameters. If this is indeed true, then we have a holomorphic Jacobi form $\frac{P}{E_4}$. This is different from other cases where the generators can be always constructed using only meromorphic modular forms with pole at the discriminant, i.e. $\tau=i \infty$. As for the bigraded ring over both weight and index, it is shown that unlike the the other cases, the weak Jacobi forms do not form a free polynomial algebra, due to an algebraic relation of some generators of low indices \cite{Wang:2018fil}. It remains a question whether the bigraded ring of weak Jacobi forms in $E_8$ case is still finitely generated. Further constructions of the basis of generators for higher index would be helpful. 

For our calculations of the elliptic genus of E-strings, the $E_8$ index is the base degree. For low base degree calculations we can use the Sakai generators for the ansatz and do not encounter the above complications. In this way we fix the elliptic genus of E-strings and M-strings up to base degree 3. In the massless limit, the Sakai generators become  the Eisenstein series, so the calculations are greatly simplified and can be pushed a little further. 

Unlike the compact Calabi-Yau model, here we can in principle fix the ansatz (\ref{EMstrings}) for all base degrees by geometric vanishing constrains, even without resorting to the help of B-model. Similarly to the unrefined case, here there are also only a finite number of non-vanishing refined Gopakumar-Vafa invariants for a given Kahler class. We again consider the restricted ansatz where we do not have non-vanishing contributions to refined Gopakumar-Vafa invariants of arbitrary high spins. Through a more sophisticated arguments than the unrefined case, we can show that the restricted ansatz is the sub-family of ansatz (\ref{EMstrings}) with only the factor $\phi_{-1, \frac{1}{2}}(\tau,  \epsilon_1) \phi_{-1,\frac{1}{2}}(\tau, \epsilon_2)$ in the denominator. It turns out the $\epsilon_{-}$ index of the numerator is negative for base degree bigger than one $\beta >1$, so the restricted ansatz must be zero. We need a simple initial non-vanishing Gopakumar-Vafa number to fix the normalization at $\beta=1$, then the ansatz (\ref{EMstrings}) is recursively completely fixed for all $\beta$. 

For the 6d SCFTs with gauge symmetry, the ansatz is more complicated than (\ref{EMstrings}), with an extra product over simple roots of the gauge algebra in the denominator, which is conjectured by comparisons with Hilbert series of the 5d theory from a circle compactification \cite{DelZotto:2016pvm}. As a result, the numerator has much bigger weight and index. We consider the cases of $n=3,4$ minimal theories, which have gauge symmetries $SU(3)$ and $SO(8)$ respectively, corresponding to $A_2$ and $D_4$ algebras. We are also content with calculations only for base degree one, which has a further simplification because we can use Jacobi forms with $SU(2)_R$ parameter $2\epsilon_{+}$ instead of $\epsilon_{+}$, due to an enhancement of symmetry. 

In these theories with gauge symmetry, the Kahler parameter of the elliptic fiber combines with the Kahler parameters of the gauge algebra, and extends its Dynkin diagram by an extra node to the affine Dynkin diagram. The modular parameter $q$ is related to the exponential of Kahler parameter $Q_0 $ of the extra node $q = Q_0\prod_{i=1}^r Q_i^{a_i^\vee}$ 
where $r$ is the rank of Dynkin diagram, $a_i^\vee$ and $Q_i$ are the comarks and corresponding exponentials of Kahler parameter of nodes. The refined Gopakumar-Vafa invariants are now labelled by the nodes of the affine Dynkin diagrams, without the modular $q$ parameter. 

\begin{figure}
\
\begin{center}
\includegraphics[angle=0,width=0.14\textwidth]{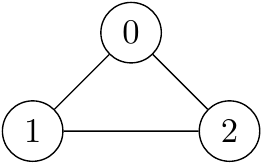} ~~~~~~
\includegraphics[angle=0,width=0.12\textwidth]{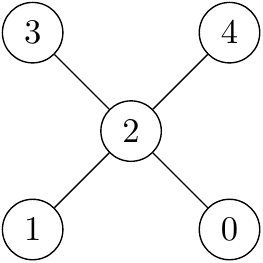}
\end{center}
\caption{Affine Dynkin diagrams for $\widehat{A}_2$ and $\widehat{D}_4$. \label{fig:dynkin}}
\end{figure}

The affine Dynkin diagrams have additional symmetry, not present in the Dynkin diagrams. For example, in Figure \ref{fig:dynkin}, we see that the affine diagram $\widehat{A}_2$ has a symmetry from permutations of 3 nodes, and the affine diagram $\widehat{D}_4$ has a symmetry from permutations of 4 nodes on the outside. The refined topological string amplitudes should have the same symmetry. It is more convenient to incorporate the symmetry into a basis of generators of affine Weyl invariant weak Jacobi forms. We construct these affine symmetric generators from Bertola's basis \cite{Bertola}, which greatly simplify the subsequent calculations.

In E-string and M-string theories, the base degree one ansatz has only one coefficient which is the overall normalization. The  {\it generic} BPS vanishing conditions give no constrain for base degree one, but are sufficient to completely fix the higher base degree ansatz. Here the generic conditions mean we only need to know that the BPS invariants always vanish at sufficiently high genus for a given Kahler class, without the precise top  genus formulas.  On the other hand, for the models with gauge symmetry, one can check that the base degree one ansatz also satisfies the generic BPS vanishing conditions, but has more coefficient than just the normalization. As before, we can fix the normalization by a simple non-vanishing invariant. In order to further fix the ansatz, we need to use the precise top genus formulas. 

From previous literature on $SU(3)$ and $SO(8)$ models \cite{Kim:2016foj, Haghighat:2014vxa}, we can compute the BPS invariants and guess the maximal genus with non-vanishing BPS numbers for a given Kahler class. There should be some geometric arguments for these vanishing conditions. Assuming the validity of these top genus formulas, we can completely fix the base degree one ansatz for the $SU(3)$ and $SO(8)$ models with the vanishing BPS conditions. 

For higher base degree, one can easily find certain sub-family of ansatz which satisfies the generic BPS vanishing conditions, due to the Lie algebra factor in the denominator of the ansatz. So unlike the E-string and M-string, the generic BPS vanishing conditions do not completely fix the higher base degree ansatz for models with gauge symmetry. We suspect that the BPS vanishing conditions with precise top genus formulas can completely fix the higher base degree ansatz for all models, although it is computationally much difficult to check than the base degree one case.

Although the results have been obtained before by path integral localization method, our weak Jacobi form ansatz method seems more efficient in some aspects, as the residue calculations in path integral become more complicated in higher base degrees. One advantage of our method is that the ansatz takes into account only the expected poles, while in the path integral calculations, the final results often seem more complicated with spurious poles, which should disappear after simplifications. Often it takes some non-trivial works involving identities of Jacobi theta functions to eliminate the spurious poles, and show the results of the two approach are indeed the same.    

\subsection{Further developments} 

It is interesting to ask whether we can study refined topological string theory on compact Calabi-Yau models. It is generally expected the refined BPS invariants can jump over the complex structure moduli space, making it difficult to define a consistent refined theory. Nevertheless, we hope to compute the refined invariants at least in some parts of the moduli space. We apply the refined holomorphic anomaly equation, which works well for non-compact models 
\cite{Krefl:2010fm, Huang:2010kf}, to the compact elliptic fibration over $\mathbb{P}^2$. Certain simple refined invariants can be computed by algebraic geometric methods. In our convention, the refined topological amplitudes from two parameter expansion are denoted $\mathcal{F}^{(n,g)}$, where $n=0$ corresponds to unrefined theory. We find that at least for the case of $n\leq 1$, the geometric computations agree with refined holomorphic anomaly equations \cite{Huangtoappear}. However, it appears for $n>1$, some modifications of the refined holomorphic anomaly equations are needed. 

We encounter two types of anomaly equation, the BCOV holomorphic anomaly equations \cite{Bershadsky:1993cx}, and the modular anomaly equations. Though it is generally expected the two type of equation are related, they have certain different features. The holomorphic anomaly equations apply for general Calabi-Yau threefolds, and appear at higher genus, while the modular anomaly equations apply only to the case of elliptic fibrations, and appear already at genus zero. Furthermore the modular anomaly equations are recursive both on genus and on the Kahler class of the base of elliptic fibration. A similar phenomenon appears in Seiberg-Witten theory, and we gave a derivation of modular anomaly equation from holomorphic anomaly equation \cite{Huang:2013eja}. We consider the connections in compact elliptic fibered Calabi-Yau threefolds \cite{Huangtoappear}. For genus zero, we derive the modular anomaly equation from Picard-Fuchs equations, based on previous works on non-compact models. For higher genus case, we derive the modular anomaly equations from BCOV holomorphic anomaly equations. The modular anomaly for BCOV propagators play important roles in the derivations.

Finally, in this section we discuss modularity in the context of writing topological string amplitudes in terms of modular forms and Jacobi forms. There is another completely different road to modularity, related to arithmetic and number theory, which has been also actively pursued. In the famous proof of Fermat's theorem by Andrew Wiles, a key ingredient is the modularity theorem, which states that elliptic curves over $\mathbb{Q}$ are related to modular forms of weight 2 of congruence subgroups. It is interesting to generalize to higher dimensional Calabi-Yau manifolds and relate to mirror symmetry. Instead of counting curves, we would now count points on the defining algebraic equations of Calabi-Yau spaces. It turns out this is indeed related to mirror symmetry and modular forms of weight 4 of congruence subgroups, in e.g. \cite{Candelas:2000fq}. The situation is clearly much richer than that of elliptic curves and deserve further investigations.

\section{Topological String Theory and Exact Quantization}   \label{section3}

Non-perturbative effects are very important in physics. Most perturbative calculations in quantum mechanics systems or quantum field theories are actually divergent asymptotic series. Of course, the underlying physical observables must be finite if we indeed have a consistent theory. The divergence in perturbative series is usually due to some non-exact mathematical manipulations, e.g. exchanging the order of performing infinite perturbative sum and (path) integral. In practice, if the expansion parameter is small, the truncation of the perturbative series at the first few terms can provide a very good approximation. Empirically the error from the true physical value is roughly the order of the last term in the truncated series, so the best approximation occurs for the optimal truncation at the minimal absolute value term of the series. In this way we have some control of the error even though the perturbative series eventually diverges. Sometimes, for example in case of QED, this is good enough for comparing with experiments. 

There are some standard mathematical ways to deal with divergent series. For example, many perturbative series in physics are Borel summable to a finite value. However, in many physically interesting cases, the Borel summation or other summation schemes still do not give the exact physical result. One has to add back non-perturbative contributions, which usually come from some non-trivial saddle points of the path integral, e.g. instanton configurations, and are non-analytic at zero coupling. There are usually some intricate connections between large order behavior of perturbative series and non-perturbative effects.

In any case, if the coupling constant of a physical system is not small, for example in the theory of QCD, we can not reliably calculate by perturbative method. Therefore it is very useful to have exact results that do not depend on perturbative method.
 
Supersymmetry has been very useful for obtaining exact results. Many supersymmetric physical observables are protected by the rigid nature of supersymmetry algebra, so are invariant under deformation of continuous parameters. This approach led to many revolutionary discoveries of dualities in quantum field theories and superstring theories, relating strong coupling physics to weak coupling physics. 

Integrable systems without necessarily supersymmetry belongs to the class of theories where exact results are also available. These systems are usually characterized by large amount of symmetries and conserved charges, and are very rare to come by. In this section we review exact results on the quantum spectrum of a class of quantum mechanics systems, derived from mirror curves of non-compact toric Calabi-Yau geometries. Some of these systems are related to known integral systems, while others have not been studied before. It turns out the exact quantization conditions can be written in terms of refined topological string amplitudes. 

Our results motivate the non-perturbative formulations of topological string theories, and probably also provide useful lessons for the more difficult case of superstring and M theories. It is generally believed that most string perturbation series are not Borel summable. In quantum field theory with a Lagrangian, we can in principle define quantum amplitudes exactly in terms of path integral for any coupling constant. This kind of non-perturbative formulations are still not available for superstring theories, which remains an outstanding fundamental problem.

\subsection{Non-relativistic quantum mechanics systems} 

In this subsection, we review some old results on conventional quantum mechanics systems, which are described by the Hamiltonian of a non-relativistic particle moving in a one-dimensional potential $H=\frac{p^2}{2m}+ V(x)$. For a quadratic potential, this is the well known solvable harmonic oscillator. However for general potential, there is no analytic solution to the Schr\"{o}dinger equation. One may treat a general potential as perturbations around the harmonic oscillator, leading to many interesting studies. 

Bender and Wu considered the perturbation of the harmonic oscillator by a quartic term $V(x) = \frac{1}{2}x^2 +g x^4$, called an anharmonic oscillator \cite{Bender:1969si}.  For $g>0$, the model is stable with discrete quantum spectrum and normalizable wave functions, while for $g<0$, the local minimum at $x=0$ is meta-stable, and the particle may decay due to quantum tunneling effects. The perturbation series for energy spectrum is factorially divergent due to the non-analyticity at $g=0$, leading to a branch cut at negative real axis. For $g>0$ the Borel summation indeed gives the exact finite quantum spectrum, while for $g<0$ there is a discontinuity of the imaginary part across the branch cut, corresponding to the decay rate. Here the decay rate is also determined by the action of the tunneling instanton, relating large order behavior of perturbation series with non-perturbative effects. 

The connections between large order behavior of perturbation series and non-perturbative effects are a universal phenomenon. In simple cases of perturbation series in differential equations or simple integrals, the relations can be explicitly derived, and is closely related to the Stokes phenomenon. In general quantum theory such relations are more difficult to prove, and are often confirmed by empirical calculations. In early 1990s, the studies of perturbation series in simple non-critical string theories showed different large order behavior from usual QFTs, and provided early hints of the nature of non-perturbative effects in string theory. 

Another type of well studied models are quantum mechanics with double-well potential. There is a nice introduction in Coleman's book \cite{Coleman}. For a symmetric potential, the perturbative energy spectra around the two local minima are degenerate. The degeneracy is broken by non-perturbative effects of order $e^{-\frac{S_0}{\hbar}}$ where $S_0$ is the action of instanton tunneling between vacua. It is not easy to find exact quantization conditions which include all instanton contributions. In early 1980s, Zinn-Justin proposed such exact quantization conditions for a class of such models \cite{ZinnJustin:1981}. The eventual proof of the quantization conditions has produced tremendous advances in modern mathematical physics, entailing the developments in the advanced theory of resurgence. The exact quantization conditions were further simplified more recently in \cite{Dunne}.  Although in this type of models, the Borel summation of perturbation series does not quite give the true physical answer,  it is shown that non-perturbative physics is still encoded in perturbation series in a more intricate way. 

Nekrasov and Shatashvili (NS) proposed that the Nekrasov partition function of $SU(N)$ Seiberg-Witten theory can describe the quantum spectra of certain integrable systems \cite{Nekrasov:2009rc}. In the correspondence, one takes a particular limit of the two Omega deformation parameters, sending one to zero while identifying the other to be the Planck constant. This is known as the NS limit. The leading term of Nekrasov free energy, namely the Seiberg-Witten prepotential $\mathcal{F}^{(0)}$, is determined by the familiar equation  $\partial_a \mathcal{F}^{(0)} =a_D$, where $a, a_D$ are the period and dual period of Seiberg-Witten curves. It turns out that Nekrasov free energy in the NS limit, which can be thought of as a quantum version of Seiberg-Witten prepotential deformed by finite Planck constant, also satisfies a similar equation, with the (dual) periods replaced by a quantum (dual) periods, which are computed from a quantum deformed Seiberg-Witten curve. This is further explored for $SU(2)$ theory (with various flavors) in \cite{MM, Huang:2012kn}. 

Nekrasov partition functions are related to topological string theory by the idea of geometric engineering \cite{Katz:1996fh}, where gauge symmetries appears from singularities of a local Calabi-Yau threefold in certain degenerate limits. In this ways we can engineer 4d and 5d gauge theories (with matter contents) from type IIA or M theory on Calabi-Yau local threefolds.  The Nekrasov partition functions for 5d gauge theories are identified with the A-model refined topological string partition functions on the corresponding Calabi-Yau threefolds, and a further limit of certain Kahler parameters reduces to the 4d case. These local Calabi-Yau geometries are described by mirror curves. By a 5d uplifting, the refined topological string free energy in NS limit is also determined by the quantum mirror maps, which are derived from a quantum deformed mirror curves \cite{Aganagic:2011mi, Huang:2012kn}. Furthermore, the exact quantization of double well quantum mechanics can be recovered by applying the NS quantization conditions to a local Calabi-Yau geometry whose mirror curve is described by the double well potential \cite{Krefl:2013bsa}.   

\subsection{Mirror curves and quantum mechanics} 

We apply the NS quantization conditions to 5d gauge theories. In this case the corresponding quantum systems are no longer that of non-relativistic particles. Instead, the quantized mirror curves correspond to quantum Hamiltonians with non-standard kinetic term, where the momentum operator appears in exponential functions. The Schr\"{o}dinger equation is no longer second order linear differential equation, but a type of more difficult difference equations. In contrast with the non-relativistic case, sometimes these type of quantum systems are called relativistic models, though there is no apparent relation with the theory of special relativity which deals with light particles moving at close to the speed of light. 

Although the quantization condition was originally proposed in the gauge theory setting, here we can further consider general Calabi-Yau geometries which do not necessarily engineer a gauge theory. For example, two commonly studied models, the local $\mathbb{P}^2$ and $\mathbb{P}^1\times \mathbb{P}^1$ models are described by the mirror curves $e^x + e^p + z e^{-x} e^{-p} =1$ and $e^x + e^p + z_1 e^{-x} +z_2 e^{-p} =1$, where the $x, p$ are complex coordinates of the curve and $z, z_1, z_2$ are the complex structure parameters. The local $\mathbb{P}^1\times \mathbb{P}^1$ model engineers the pure $SU(2)$ supersymmetric gauge theory in 5d, while local $\mathbb{P}^2$ model has no immediate gauge theory counterpart. We promote the $x, p$ coordinate to the quantum canonical position and momentum operators satisfying the usual commutation relation $[\hat{x}, \hat{p}]=i\hbar$, and the complex structure parameter is identified with the exponential of Hamiltonian. For simplicity we take $z_1=z_2$ in local $\mathbb{P}^1\times \mathbb{P}^1$ model, although the general case can be also studied. In this way we derive the quantum Hamiltonians 
\begin{eqnarray} \label{models1}
&& e^{\hat{H} }=  e^{\hat{x}} + e^{\hat{p}} + e^{ -\hat{x}-\hat{p}} , ~~~~\textrm{local}~ \mathbb{P}^2~\textrm{model},   \\ 
&& e^{\hat{H}} =  e^{\hat{x}}+ e^{-\hat{x}}  + e^{\hat{p}} + e^{-\hat{p}} , ~~\textrm{local}~ \mathbb{P}^1\times \mathbb{P}^1~\textrm{model}.  \nonumber 
\end{eqnarray} 
The Hamiltonian of the local $\mathbb{P}^1\times \mathbb{P}^1$ model appeared also in the studies of integrable systems as the relativistic Toda model. 

These Hamiltonians are apparently bounded below, have infinite energy in the asymptotic region of phase space $(x,p)$, therefore should have discrete quantum spectra and normalizable wave eigenfunctions. Although there is no known analytic solution, it is possible to compute the quantum spectra numerically using the basis of eigenfunctions of a harmonic oscillator. The matrix element of the Hamiltonian with two eigenstates of a harmonic oscillator can be computed analytically. We can truncate the energy level to a finite number, and diagonalize a finite matrix to find the energy eigenvalues of the Hamiltonians in (\ref{models1}). As we take the truncation level to be large, in principle we can compute the energy eigenvalues to any numerical precision. 

It turns out that in contrast with the non-relativistic models, there are some new non-perturbative contributions, first noticed in \cite{Kallen:2013}.  One encounters an immediate difficulty for simply applying the original 4d NS quantization condition as in non-relativistic case. The refined topological string amplitudes in NS limit have poles when the Planck constant $\hbar$ is a rational number times $\pi$, so naively they can not give the correct quantization conditions.  On the other hand, the Hamiltonians like (\ref{models1}) appear to be consistent quantum systems and there seems no apparent mechanism for potential pathology at these densely discrete values of Planck constant. Furthermore, the 4d NS quantization condition is analytic around $\hbar\sim 0$, it can be checked that it agrees with the perturbative expansion of the Hamiltonians around $\hbar\sim 0$. So the only plausible resolution of the apparent difficulty is to postulate some new non-perturbative contributions which are not analytic around $\hbar\sim 0$. Motivated by the studies of M2 brane theories, Kallen and Mari\~{n}o  propose such non-perturbative contributions for the local $\mathbb{P}^1\times \mathbb{P}^1$ model \cite{Kallen:2013}. Their non-perturbative contributions are constructed from unrefined topological string amplitudes and intricately cancel all poles of the Planck constant. It should be noted that the singularity cancellations alone do not completely fix the non-perturbative contributions. One could write many other additional non-perturbative contributions which are smooth for $h>0$ so do not interfere with singularity cancellations. Some numerical tests are performed and seem to support the proposal within uncertainty of numerical precision. 

However it further turns out that the proposal in \cite{Kallen:2013} is still not complete. We study the proposal in more models and also perform more precise numerical tests of the quantum spectra \cite{Huang:2014eha}.  These more precise tests reveal very tiny but irreconcilable difference of the quantum spectra computed from the harmonic oscillator basis and from the proposed exact quantization condition. To remedy the situation, we compute the discrepancies for various values of Planck constant, and luckily they can be numerically fitted to some simple smooth functions. We determine some of these smooth corrections in low orders that could be added to the non-perturbative contributions to account for the discrepancies. Of course, this numerical fitting approach is quite tedious, and without a guiding principle there is no hope to completely determine an apparent infinite series of smooth corrections with decreasing magnitudes. 

Grassi, Hatsuda and Mari\~{n}o improve upon the previous works and propose a formula for the spectral determinant of the Hamiltonian, which we shall called the GHM conjecture  \cite{Grassi:2014zfa}.  The formula uses the NS limit of refined topological string amplitudes as the perturbative part and still the unrefined limit as the non-perturbative part. The spectral determinant is defined as $\Theta(E) = \prod_{n} (1-e^{E-E_n})$, as such the zero points of the energy $E$ are exactly the eigenvalues. Setting the  spectral determinant to zero thus gives an exact quantization condition. In this way, one can analytically derive the infinite series of smooth non-perturbative contributions, and the low oder terms agree with previous results from numerical fittings. This nice development appears to bring a satisfactory completion to the quests for exact quantization condition for these models. 

In yet another surprising twist of the story, we later propose a different exact quantization condition \cite{Wang:2015wdy}. In the Bohr-Sommerfeld form, we have 
\begin{eqnarray}  \label{BScondition}
\textrm{vol}_{\rm p}(E, \hbar) + \textrm{vol}_{\rm np}(E, \hbar) = 2\pi \hbar(n+\frac{1}{2}), 
\end{eqnarray}
where $n=0,1,2, \cdots$ is the discrete energy level, and the formulas for perturbative and non-perturbative parts 
$\textrm{vol}_{\rm p}(E, \hbar)$ and $\textrm{vol}_{\rm np}(E, \hbar)$ are 
\begin{eqnarray} \label{vol}
\textrm{vol}_{\rm p}(E,\hbar ) &=& \frac{\tilde{t}^2-\pi^2}{2} -\frac{\hbar^2}{8}+\hbar   f_{NS} (\tilde{t} , \hbar), 
 \nonumber \\
\textrm{vol}_{\rm np}(E,\hbar ) &=& \hbar   f_{NS} (\frac{2\pi \tilde{t}}{\hbar} , \frac{4\pi^2}{\hbar}). 
\end{eqnarray}
Here the quantum mirror map $\tilde{t}(E)$ is a function of the energy $E$ which is computed from the quantum mirror curve. The function $f_{NS} (\tilde{t} , \hbar)$ is the NS limit of refined topological string amplitudes 
\begin{eqnarray} \label{fNS}
f_{NS} (\tilde{t} , \hbar) &=& - \frac{3}{2} \sum_{j_L,j_R} \sum_{w,d=1}^{\infty}   (-1)^{2j_L+2j_R+rwd}   n^{d}_{j_L,j_R}    e^{wd\tilde{t}}   \nonumber \\ & \times & \frac{d}{w} \cdot
\frac{\sin \frac{w\hbar(2j_R+1)}{2} \sin \frac{w\hbar(2j_L+1)}{2}}{ \sin^3 (\frac{w\hbar}{2})}, 
\end{eqnarray}
where $n^{d}_{j_L,j_R}$ is the refined Gopakumar-Vafa invariant and $r$ is an integer depending on models, later referred to as ``r-field".  

We see that the non-perturbative contributions can be simply obtained by the replacement of variables 
\begin{equation} \label{transform}
\tilde{t}\rightarrow \frac{2\pi \tilde{t}}{\hbar}, ~~~~ \hbar\rightarrow \frac{4\pi^2}{\hbar},
\end{equation}
in the $f_{NS} (\tilde{t} , \hbar)$  part of the perturbative contributions. This is tantalizingly similar to the S-duality transformation well known in certain quantum field theories. Because our quantization uses the NS limit for both perturbative and non-perturbative contributions, we often still called it the NS quantization condition, to distinguish it from the GHM quantization condition. This should not be confused with NS quantization condition for non-relativistic models from 4d gauge theory, which essentially requires no non-perturbative parts.

Our conjecture is partly inspired by earlier works on non-perturbative completion of topological string theory \cite{Lockhart:2012}, but the specific motivations and contexts are different. For example, they are mainly motivated by a $SL(3,\mathbb{Z})$ symmetry, and are not in particular looking for quantization conditions so are not using the quantum mirror map. In the context of GHM conjecture, Mari\~{n}o and collaborators also propose to non-perturbatively formulate the topological strings as matrix models derived from the quantum Hamiltonian operators \cite{Grassi:2014zfa}.  Some other proposals for non-perturbative (refined) topological string theory have appeared in the past, with different perspectives and virtues.  For example, there is a nice proposal to determine the non-perturbative transseries in topological string theory by a generalization of BCOV holomorphic anomaly equations \cite{Santamaria:2013rua}.  As a general story, a non-perturbative proposal usually emerges for a theory previously only defined perturbatively, due to its conjectural equivalence with another non-perturbatively defined theory. Here our quantum mechanics systems are well defined non-perturbatively for any $\hbar>0$, and provide leverage for a non-perturbative formulation of the corresponding topological string theory. It is always interesting if one can perform tests of the predictions of a non-perturbative formulation away from the weak coupling regions.

Although our conjectured quantization condition looks quite different from the GHM conjecture, we can check the non-perturbative parts agree at some low orders. We conjecture they are indeed equivalent, which impose some non-trivial relations among the refined Gopakumar-Vafa invariants. Later it is shown in \cite{Grassi:2016nnt}  that the equivalence can be derived from a set of differential equations for the Nekrasov's partition function, known as the blowup equations, first appeared in the study of instanton partition functions on the moduli spaces on the blowup of $\mathbb{R}^4$ \cite{Nakajima:2003pg}. 

Out main examples here are mirror curves of genus one. The correspondence of topological string theory and quantum systems can be generalized to the case of higher genus mirror curves. In these cases, one has multiple commuting Hamiltonians and quantization conditions. The correspondence provides new exact quantization conditions for many integrable systems, in e.g. \cite{Hatsuda:2015qzx}. We also study the equivalence of NS and GHM quantization conditions for the cases of higher genus mirror curves and propose a general procedure to determine  the r-fields  \cite{Sun:2016obh}. Of course, it would be far more interesting if the correspondence works for compact Calabi-Yau geometries. The main obstruction is that the compact models are no longer described by the (complex) one-dimensional mirror curves, but are intrinsically higher dimensional.  

Motivated by the appearance of blowup equations, we turn around the logic and use the blowup equations to compute the topological string  amplitudes \cite{Huang:2017mis}. The blowup equations are originally derived for instanton partition functions of 4d and 5d supersymmetric gauge theories. However, in many cases, similar blowup equations appear for refined topological string amplitudes on non-compact toric Calabi-Yau geometries even they do not correspond to any gauge theory through geometric engineering. In most of these cases, there is no mathematical proofs of the blowup equations at the moment. We find that in most cases, the blowup equations can completely determine the refined topological string amplitudes. This provides another new way for computation even these models can be solved previously by refined topological vertex and BCOV methods. More recently, the blowup equations are applied to more complicated models of non-toric elliptic Calabi-Yau corresponding to minimal 6d SCFTs \cite{Gu:2018gmy}, discussed in Section \ref{section2}.

The quantum mechanics systems described here are relevant for observable experiments in condensed matter physics. For example, the local $\mathbb{P}^1\times \mathbb{P}^1$ model is related to the rather famous Hofstadter's butterfly, a fractal structure in the energy spectrum of electrons moving in a 2d periodic potential under a uniform magnetic field. The pattern was predicted by Hofstadter in 1976 and eventually experimentally observed many years later. Here our quantum mechanics systems have only bound states with normalizable wave functions. On the other hand, in Hofstadter's model, the electrons are not localized, but have periodic two-dimensional un-normalizable wave functions. Nevertheless, the mathematics of the two different physical theories are intimately related \cite{Hatsuda:2016mdw, Hatsuda:2017zwn}. Many other examples related to integrable systems are explored in the literature.

\section{Conclusion}
It is clear that there are many interesting deep problems in the field remained to be explore, which should be sufficient to support decades of more research. We think there will be more fundamental discoveries in the future. Meanwhile, it is always interesting to find connections of topological string theory with other topics in mathematics and theoretical physics.

\

\noindent \textit{Acknowledgements}: We thank our collaborators of the research works reviewed in this article. This work was supported by the National Natural Science Foundation of China (Grant No. 11675167) and the national ``Young Thousand People" program.

\end{document}